\documentclass[prd,aps,twocolumn,showpacs,dvips]{revtex4}%
\usepackage{feynmp}
\usepackage{amssymb}
\usepackage{epsfig}
\usepackage{graphicx}
\usepackage{subfigure}
\usepackage{hyperref}

\begin{document}
\title{Dynamical chiral symmetry breaking in QED$_{3}$ at finite density and impurity potential}
\author{Wei Li and Guo-Zhu Liu \\
{\small {\it Department of Modern Physics, University of Science and
Technology of China, Hefei, Anhui, 230026, P.R. China }}}

\begin{abstract}
We study the effects of finite chemical potential and impurity
scattering on dynamical fermion mass generation in (2+1)-dimensional
quantum electrodynamics. In any realistic systems, these effects
usually can not be neglected. The longitudinal component of gauge
field develops a finite static length produced by chemical potential
and impurity scattering, while the transverse component remains
long-ranged because of the gauge invariance. Another important
consequence of impurity scattering is that the fermions have a
finite damping rate, which reduces their lifetime staying in a
definite quantum state. By solving the Dyson-Schwinger equation for
fermion mass function, it is found that these effects lead to strong
suppression of the critical fermion flavor $N_c$ and the dynamical
fermion mass in the symmetry broken phase.
\end{abstract}


\maketitle


\section{Introduction}

The dynamical chiral symmetry breaking (DCSB) in (2+1)-dimensional
quantum electrodynamics (QED$_{3}$) has been investigated
intensively for more than twenty years \cite{Pisarski, Appelquist86,
Appelquist88, Nash, Dagotto, Atkinson90, Aitchison92, Pennington,
Maris, Gusynin96, Fischer, Hands, Gusynin03, Roberts}. On the one
hand, these investigations may help to gain deeper understanding of
DCSB in QCD. On the other hand, this non-perturbative phenomenon
gives an elegant field theoretic description for the formation of
long-range antiferromagnetic order in two-dimensional quantum
Heisenberg antiferromagnet \cite{AFM, Liu02, Liu03}.

Using the Dyson-Schwinger (DS) equation approach, Appelquist
\emph{et} \emph{al.} first found that DCSB caused by fermion mass
generation takes place only when the fermion flavor $N$ is below
some critical value $N_{c}$ \cite{Appelquist88}. When $N > N_{c}$,
the fermions remain massless and the chiral symmetry is preserved.
Although there was some controversy about the existence and precise
value of $N_{c}$ in the literature, most analytical and numerical
computations confirmed that $N_{c} \approx 3.5$ \cite{Appelquist88,
Nash, Dagotto, Maris, Gusynin96, Fischer, Roberts}.

The DCSB is realized by forming fermion-anti-fermion pairs mediated
by strong gauge interaction. In order to trigger this low-energy
phenomenon, an essential requirement is that the gauge interaction
between fermions should be sufficiently long-ranged. If the gauge
field develops an effective screening length, then the gauge
interaction is significantly weakened. We have considered the effect
of a finite gauge boson mass $m_{a}$ on DCSB previously
\cite{Liu03}, and found that the increasing $m_{a}$ leads to a
significant suppression of DCSB.

When the action contains only massless Dirac fermions and
non-compact gauge field, the gauge interaction is always long-ranged
since the gauge invariance ensures that no photon mass term can be
generated by the radial correction from fermions. If the gauge field
also couples to an additional Abelian Higgs model, then the photon
can acquire a finite mass via the Anderson-Higgs mechanism in the
superconducting state \cite{Liu03}. In the context of high
temperature superconductor, the suppression of DCSB by photon mass
$m_{a}$ can be used to qualitatively understand the competition
between antiferromagnetism and superconductivity, which is one of
the most prominent phenomena in high temperature superconductors
\cite{Liu03}.

Apart from Anderson-Higgs mechanism, the gauge field may also be
screened by other physical effects. For example, when the fermion
density is nonzero, the chemical potential $\mu$ should be
explicitly taken into account. In addition, the fermions may be
scattered by some impurity atom (defect or imperfection) in
realistic interacting system described by QED$_3$. Although
generally chemical potential and impurity scattering have distinct
influences on the physical properties of the interacting fermion
system, there is a common feature that they can both induce a finite
density of states at the Fermi level. This screens the time
component of the gauge field, which then becomes short-ranged.
However, the transverse gauge field remains long-ranged in the sense
that the static screening length still diverges as at zero chemical
potential in the clean limit ("clean" means there is completely no
impurity scattering). In principle, the long-range nature of
transverse gauge field is guaranteed by the gauge invariance.
However, though the transverse part is unscreened, the effective
gauge interaction strength is reduced by the fermion density caused
by chemical potential and/or impurity scattering process, which
certainly affects the critical flavor $N_{c}$ as well as the
dynamical fermion mass. Besides generating finite fermion density,
the impurity scattering leads to the damping of fermionic quantum
states by producing a finite scattering rate $\Gamma_0$. Such
damping effect reduces the lifetime of a fermion staying at a
definite quantum state, specified by such quantum numbers as
momentum and/or energy. This will also have important effects on
DCSB.

While there appeared several papers discussing the effect of
chemical potential on DCSB in QED$_{3}$ \cite{Feng}, to our
knowledge the effect of impurity scattering has never been
considered in previous work. Indeed, in any realistic applications
of QED$_{3}$ to condensed matter physics, such as high temperature
superconductor, the chemical potential and impurity scattering are
usually important and thus should not been ignored. The purpose of
this paper is to study the dependence of DCSB on finite density and
impurity scattering.

The paper is organized as follows. In Sec. \ref{sec:model}, we set
up the Lagrangian and then discuss the screening effect of gauge
interaction by including chemical potential and impurity scattering
rate into the vacuum polarization function. We then solve the
Dyson-Schwinger equation and present the critical flavor and fermion
mass function in Sec. \ref{sec:solution}. The discussion is
presented in Sec. \ref{sec:summary}.

\section{Dyson-Schwinger equation in the presence of $\mu$ and $\Gamma_0$} \label{sec:model}

The Lagrangian for (2+1)-dimensional QED with $N$ flavors of
massless fermions is given by
\begin{equation}\label{eq:L_qed3}
{\cal L} = - \frac{1}{4}F_{\mu\nu}F^{\mu\nu}
+ \sum_{a=1}^{N} \overline{\psi}_{a}(i\partial\!\!\!\slash_{\mu} - eA\!\!\!\slash_{\mu})\psi_{a}.
\end{equation}
In (2+1)-dimensional space-time, the lowest rank spinorial
representation is two-component spinor whose $2 \times 2$
representation may be chosen as the Pauli matrices $\gamma_{\mu} =
(\sigma_2,i\sigma_3,i\sigma_1)$. However, it is impossible to define
a $2 \times 2$ matrix that anticommutes with all these matrices.
Therefore, there is no chiral symmetry in this representation. The
fermion can be described by a four-component spinor field $\psi$,
whose conjugate spinor field being defined as $\bar{\psi} =
\psi^{\dagger}\gamma_{0}$ \cite{Appelquist86}. The $4\times 4$
$\gamma$-matrices can be defined as $\gamma_{\mu} =
(\sigma_3,i\sigma_1,i\sigma_2)\otimes\sigma_{3}$, satisfying the
standard Clifford algebra $\{\gamma_{\mu}, \gamma_{\nu}\} =
2g_{\mu\nu}$ with metric $g_{\mu\nu} = \mathrm{diag}(1,-1,-1)$.  It
is easy to verify that there are two $4\times 4$ matrices
\begin{eqnarray}
  \gamma_{3} = i\left(
  \begin{array}{cc}
    0 & I \\
    I & 0 \\
  \end{array}
\right)
,\,\,\,\,\,\,\
    \gamma_{5} = i\left(
  \begin{array}{cc}
    0 & I \\
    -I & 0 \\
  \end{array}
\right),\nonumber
\end{eqnarray}
which anticommute with all $\gamma_{\mu}$. The massless Lagrangian
(Eq. (\ref{eq:L_qed3})) preserves a continuous U(2N) chiral symmetry
$\psi \rightarrow e^{i\alpha\gamma_{3,5}}\psi$. The mass term
generated by fermion-anti-fermion pairing will break this global
chiral symmetry dynamically to subgroup $U(N)\times U(N)$. In the
following we consider a general large $N$ and perform the $1/N$
expansion. For convenience, we work in units where $\hbar = k_{B} =
1$.

In the Euclidian space, the full gauge field propagator
$D_{\mu\nu}(q)$ is given by the equation
\begin{equation}\label{eq:photon_full_0}
D_{\mu\nu}^{-1}(q) = D_{\mu\nu}^{(0) - 1}(q) + \Pi_{\mu\nu}(q),
\end{equation}
with the free photon propagator being
\begin{equation}\label{eq:photon_0}
D_{\mu\nu}^{(0)}(q) = \frac{1}{q^{2}}(g_{\mu\nu} - \frac{q_{\mu}q_{\nu}}{q^2}),
\end{equation}
in the Landau gauge. To the leading order of $1/N$ expansion, the
one-loop contribution to vacuum polarization tensor $\Pi_{\mu\nu}$
is
\begin{eqnarray}\label{eq:pi}
\Pi_{\mu\nu}(q) = - \alpha\int \frac{d^{3}k}{(2\pi)^{3}}
\frac{\mathrm{Tr}[\gamma_{\mu}k\!\!\!/\gamma_{\nu}(q\!\!\!/ + k\!\!\!/)]}{k^2(q+k)^2},
\end{eqnarray}
where $\alpha=Ne^2$. The QED$_{3}$ can be treated using the $1/N$
expansion, with the product $\alpha = Ne^2$ being fixed when $N
\rightarrow \infty$ and $e^{2} \rightarrow 0$. This tensor can also
be written as $\Pi_{\mu\nu}(q) =
(g_{\mu\nu}-\frac{q_{\mu}q_{\nu}}{q^2})\Pi(q^2)$ according to the
gauge invariance. It is easy to find the polarization function:
$\Pi(q)=\frac{\alpha q}{8}$. Now the propagator of gauge field has
the form
\begin{equation}\label{eq:photon0}
D_{\mu\nu}^{-1}(q) = \frac{1}{q^{2} + \Pi(q)}(g_{\mu\nu} - \frac{q_{\mu}q_{\nu}}{q^2}).
\end{equation}

In order to study the dynamical fermion mass generation, one can
write the following DS equation for fermions propagator
\begin{equation}\label{eq:dsf0}
S_{F}^{-1}(p)=S_{F}^{(0)-1}(p) - e^{2} \int\frac{d^{3}k}{(2\pi)^{3}}
\gamma_{\mu}S_{F}(k)D_{\mu\nu}(p-k)\Gamma_{\nu}(k,p).
\end{equation}
To the leading order in $1/N$ expansion, the vertex function
$\Gamma_{\nu}$ is replaced by the bare matrix $\gamma_{\nu}$. The
inverse full propagator of fermion is
\begin{equation}
S_{F}(p)^{-1}=ip\!\!\!/ + \Sigma(p),
\end{equation}
where the wave-function renormalization is neglected. Taking trace
on both sides of the DS equation, we get an integral equation
\begin{equation}
\Sigma(p) = \frac{e^2}{4}\int\frac{d^{3}k}{(2\pi)^{3}}
\frac{\Sigma(k)}{k^2+ \Sigma^2(k)}
\mathrm{Tr}[\gamma_{\mu}D_{\mu\nu}(p-k)\gamma_{\nu}].
\end{equation}
Using the gauge field propagator Eq. (\ref{eq:photon0}), we have
\begin{equation}\label{eq:mass0}
\Sigma(p)=\frac{2\alpha}{N}\int\frac{d^{3}k}{(2\pi)^{3}}
\frac{\Sigma(k)}{k^2 + \Sigma^2(k)}\frac{1}{(p-k)^2 + \frac{\alpha (p-k)}{8}},
\end{equation}
which was first obtained by Appelquist \emph{et} \emph{al.}
\cite{Appelquist88}. They found that DCSB takes place only when the
fermion flavor is less than a critical value $N_{c} = 32/\pi^{2}$.
Motivated by this interesting prediction, a great many attentions have
been paid on this issue. After taking into account the high order
corrections to wave function renormalization, the critical flavor
was found to change to $N_{c} = 128/3\pi^{2}$ \cite{Nash}.
Pennington \emph{et} \emph{al} \cite{Pennington} used a more careful
truncation of the fermion DS equation and found that DCSB can occur
for all values of $N$. When the DS equations of fermion self-energy
is coupled self-consistently to those of gauge field propagator,
Maris \cite{Maris} showed that $N_{c} \simeq 3.3$. In a recent
publication, Fischer \emph{et} \emph{al.} \cite{Fischer} found that
$N_{c} \simeq 4$ after detailed analysis of vertex corrections. Some
numerical simulations on lattice QED$_{3}$ claim that there is no
DCSB for $N > 2$ \cite{Hands}. However, Gusynin \emph{et} \emph{al.}
found that the absence of DCSB can be attributed to the large
infrared cutoff used in lattice studies and the smallness of the
generated mass scale \cite{Gusynin03}. In summary, most analytical
and numerical computations seem to agree that the critical fermion
flavor should be $N_c \approx 3.5$ \cite{Appelquist88, Nash,
Dagotto, Maris, Gusynin96, Fischer, Roberts}, close to the original
value obtained by Appelquist \emph{et} \emph{al.} within the lowest
order of $1/N$ expansion.

The above investigation and result are valid at zero chemical
potential in clean fermion system. Till now, little attention has
been paid to the case with finite fermion density and finite
impurity potential. Since the chemical potential and impurity
scattering are generally very important in realistic applications,
it is important to examine their effects on DCSB. The aim of this
paper is to study this problem.

First of all, at nonzero chemical potential the fermions have a
finite density at the Fermi level, which screens the temporal
component of gauge field and weakens the gauge interaction. The
influence of chemical potential on DCSB was previously discussed by
Feng \emph{et} \emph{al.} \cite{Feng}. They derived an explicit
equation to include the chemical potential into the DS equation. One
crucial assumption in their work is the neglecting of chemical
potential in the vacuum polarization function. The numerical results
of Feng \emph{et} \emph{al.} show that the chemical potential leads
to only insignificant change of $N_c$ and fermion mass. We think
that the screening effect caused by chemical potential is very
important and hence will pay special attention to the screening of
gauge field induced by chemical potential by studying the vacuum
polarization. As will be shown in the context, the chemical
potential suppresses both $N_c$ and fermion mass strongly.

The effect of impurity scattering is more complicated than chemical
potential. Generally, the scattering of fermions by impurity
potential has two important effects. First, it generates a finite
density of states at low energy, which also screens the gauge field.
This effect is expected to weaken the gauge interaction, analogous
to the role played by the chemical potential. Second, the impurity
scattering produces a finite damping of fermion quantum states and
thus reduces the time for massless Dirac fermions to interact with
their anti-particles. These two effects are both very important.

The behavior of massless Dirac fermions in a random potential is of
great interest in the context of condensed matter physics since the
low-energy properties of some many-body systems, such as $d$-wave
high temperature superconductor and graphene, are largely controlled
by the interaction of Dirac fermion with impurities. Unfortunately,
at present the problem of random Dirac fermion has not been fully
understood even when there is no direct interaction between
fermions. When the impurity scattering and gauge interaction are
both important, the problem is basically out of theoretical control.
If we consider a single impurity atom, then the impurity scattering
can be treated by the self-consistent Born approximation. Within
this approximation, the retarded fermion self-energy function
develops a finite imaginary part, which is usually represented by a
constant scattering rate $\Gamma_{0}$. To study the problem about
impurity scattering, it is most convenient to work in the Matsubara
formalism and write the fermion propagator as
\begin{eqnarray}\label{eq:propagator_Gamma}
S_{F}(i\omega_n,\mathbf{p}) = \frac{1}{i\omega_n\gamma_{0} -
\mathbf{\gamma}\cdot \mathbf{p}},
\end{eqnarray}
where the frequency is $i\omega_n = i\frac{(2n+1)\pi}{\beta}$ with
$\beta = \frac{1}{T}$. Once the impurity scattering rate is taken
into account, the fermion frequency should be replaced by
\begin{equation}
i\omega_{n} \rightarrow i\omega_{n} + i\Gamma_{0}\mathrm{sgn}(\omega_{n}).
\end{equation}
The fermion damping effect can be intuitively understood as follows.
After analytical continuation,
\begin{equation}
i\omega_{n} \rightarrow \omega + i\delta,
\end{equation}
the retarded fermion propagator has the form
\begin{eqnarray}\label{eq:propagator_Gamma}
S_{F}^{\mathrm{ret}}(\omega,\mathbf{p}) = \frac{1}{(\omega + i\Gamma_{0})\gamma_{0} -
\mathbf{\gamma}\cdot \mathbf{p}}.
\end{eqnarray}
After Fourier transformation, it becomes
\begin{equation}
S_{F}(t,\mathbf{r}) = e^{i\omega t - \Gamma_{0}t}e^{i\mathbf{p}\cdot
\mathbf{r}}
\end{equation}
in real space. Apparently, the parameter $\Gamma_{0}$ measures the
decaying rate of the fermionic state characterized by such quantum
numbers as $(\omega,\mathbf{p})$, which is known as the Landau
damping effect. Starting from the propagator Eq.
(\ref{eq:propagator_Gamma}), various physical quantities can be
calculated and compared with experiments. Intuitively, the damping
effect is at variance with DCSB since a fermion may be scattered
into another state before it combines with its anti-particle to form
a stable pair.

Before we set up the DS equation for fermion self-energy, we need
first to calculate the photon propagator and discuss the screening
effect caused by chemical potential and impurity scattering within
the Matsubara formalism. With finite chemical potential $\mu$ and
finite damping rate $\Gamma_{0}$, the fermion propagator reads
\begin{eqnarray}\label{eq:propagator_full}
S_{F}(i\omega_{n},\mathbf{p}) = \frac{1}{(i\omega_{n} +
i\Gamma_{0}\mathrm{sgn}(\omega_{n})-\mu)\gamma_{0} -
\mathbf{\gamma}\cdot \mathbf{p}}.
\end{eqnarray}
The energy shift caused by chemical potential and the damping effect
caused by impurity scattering are both reflected in this propagator.
The screening of gauge interaction induced by them can be seen from
the corresponding vacuum polarization functions. We will use this
propagator to calculate the vacuum polarization functions and to
construct the DS equation for fermion mass.

The inverse photon propagator for frequency $q_{0} =
\frac{2m\pi}{\beta}$ and spatial momentum $|{\bf q}|$ is given by
\begin{equation}\label{eq:photon_full_t}
\Delta_{\mu\nu}^{-1}(q_{0},{\bf q}, \beta) = \Delta_{\mu\nu}^{(0)
-1}(q_{0},{\bf q}, \beta) + \Pi_{\mu\nu}(q_{0},{\bf q},\beta),
\end{equation}
where the $\Delta_{\mu\nu}^{(0)}$ is the free photon propagator.
Here, we use different symbols $D_{\mu\nu}$ and $\Delta_{\mu\nu}$ to
denote the gauge boson propagator at zero temperature and finite
temperature, respectively. Taking advantage of the transverse
condition, $q_{\mu}\Pi_{\mu\nu}(q) = 0$, the vacuum polarization
tensor defined by Eq. (\ref{eq:pi}) can be decomposed in terms of
two independent transverse tensors \cite{Dorey},
\begin{eqnarray}
\Pi_{\mu\nu}(q_{0},\mathbf{q},\beta)
= \Pi_{A}(q_{0},\mathbf{q},\beta) A_{\mu\nu}
+ \Pi_{B}(q_{0},\mathbf{q},\beta) B_{\mu\nu},
\end{eqnarray}
where
\begin{eqnarray}
A_{\mu\nu} &=& \Big(\delta_{\mu 0} -
\frac{q_{\mu}q_{0}}{q^{2}}\Big)\frac{q^{2}}{\mathbf{q}^{2}}
\Big(\delta_{0\nu} - \frac{q_{0}q_{\nu}}{q^{2}}\Big), \\
B_{\mu\nu} &=& \delta_{\mu i}\Big(\delta_{ij} -
\frac{q_{i}q_{j}}{\mathbf{q}^{2}}\Big)\delta_{j\nu}.
\end{eqnarray}
They are orthogonal and related by the relationship
\begin{equation}
A_{\mu\nu} + B_{\mu\nu} = \delta_{\mu\nu} - \frac{q_{\mu}q_{\nu}}{q^{2}}.
\end{equation}
The functions $\Pi_{A}(q_{0},\mathbf{q},\beta)$ and $\Pi_{B}(q_{0},\mathbf{q},\beta)$ are related to the temporal and spatial components of vacuum polarization tensor $\Pi_{\mu\nu}$ by the following expressions
\begin{equation}
\Pi_{A} = \frac{q^{2}}{\mathbf{q}^{2}}\Pi_{00}, \,\,\,\,\,\
\Pi_{B} = \Pi_{ii} - \frac{q_{0}^{2}}{\mathbf{q}^{2}}\Pi_{00}.
\end{equation}
Now the full finite temperature photon propagator $\Delta_{\mu\nu}(q_{0}, {\bf q}, \beta)$ can be written as
\begin{equation}\label{eq:photon_full}
\Delta_{\mu\nu}(q_{0}, {\bf q}, \beta) =
\frac{A_{\mu\nu}}{q^{2} + \Pi_{A}(q_{0}, {\bf q}, \beta)}
+ \frac{B_{\mu\nu}}{q^{2} + \Pi_{B}(q_{0}, {\bf q}, \beta)}.
\end{equation}
At finite temperature, the fermion contribution to vacuum polarization functions should be
\begin{widetext}
\begin{eqnarray}\label{eq:pi_full}
\Pi_{\mu\nu}(q,\beta) &=&
- \frac{\alpha}{\beta}\sum_{n = -\infty}^{\infty}\int\frac{d^{2}\mathbf{K}}{(2\pi)^{2}}
\frac{\mathrm{Tr}[\gamma_{\mu}k\!\!\!/\gamma_{\nu}(q\!\!\!/+k\!\!\!/)]}{k^2(q+k)^2}
\nonumber \\ &=&
-\frac{4\alpha}{\beta}\int_{0}^{1}dx\sum_{n=-\infty}^{\infty}
\int\frac{d^{2}\mathbf{L}}{(2\pi)^{2}}\frac{2l_{\mu}l_{\nu} +
(1-2x)(l_{\mu}q_{\nu}+q_{\mu}l_{\nu} - q\cdot l\delta_{\mu\nu}) +
2x(1-x)(q^{2}\delta_{\mu\nu}-q_{\mu}q_{\nu})}{\left[l^{2}+x(1-x)q^{2}\right]^{2}}
\nonumber \\ &&
-\frac{4\alpha}{\beta}\int_{0}^{1}dx\sum_{n=-\infty}^{\infty}
\int\frac{d^{2}\mathbf{L}}{(2\pi)^{2}}\frac{-\delta_{\mu\nu}}{l^{2}+ x(1-x)q^{2}},
\end{eqnarray}
\end{widetext}
where $q=(q_0,\mathbf{q})$, $Q=|\mathbf{q}|$, $q_0=\frac{2m\pi}{\beta}$ for gauge boson and $k =(k_0,\mathbf{k})$, $K=|\mathbf{k}|$, $k_0=\frac{(2n+1)\pi}{\beta}$ for fermion. Here, a new momentum variable is defined by $l = k + xq$ with
\begin{equation}
l=(l_{0},\mathbf{l}),\,\,\, L=|\mathbf{l}|,\,\,\,
l_{0}=\frac{2\pi}{\beta}(n+xm+\frac{1}{2}).
\end{equation}
In the presence of impurity scattering rate $\Gamma_{0}$ and chemical potential $\mu$, the variable $l_{0}$ should be replaced by
\begin{equation}
l_{0}(n) = \frac{2\pi}{\beta}(n + xm + \frac{1}{2} + i\frac{\beta}{2\pi}\mu +
\frac{\beta}{2\pi}\Gamma_{0}\mathrm{sgn}(\omega_{n})).
\end{equation}
Now the spatial and temporal component of polarization tensor $\Pi_{\mu\nu}$ can be expressed as
\begin{eqnarray}\label{eq:pi_ij}
&&\Pi_{ij}(q_0,\mathbf{q},\beta) = \frac{4\alpha}{\beta} \int_{0}^{1}dx
\int\frac{d^{2}\mathbf{L}}{(2\pi)^{2}}
\nonumber \\&&\,
\times[2x(1-x)(q^{2}\delta_{ij}-q_{i}q_{j})S_{2} - (1-2x)\delta_{ij}q_{0}S^{*}],
\end{eqnarray}
\begin{eqnarray}\label{eq:pi_00}
&&\Pi_{00}(q_0,\mathbf{q},\beta) = \frac{4\alpha}{\beta} \int_{0}^{1}dx
\int\frac{d^{2}\mathbf{L}}{(2\pi)^{2}}
\nonumber \\&&\,
\times\left[S_{1} - 2(L^{2}+x(1-x)q_{0}^{2})S_{2} + (1-2x)q_{0}S^{*}\right].
\end{eqnarray}
where, following Ref. \cite{Dorey}, we defined
\begin{eqnarray}
S_{i} &=& \sum_{n=-\infty}^{\infty}\frac{1}{\left[l_{0}^{2}(n) + L^{2}
+ x(1-x)q^{2}\right]^{i}}, \\
S^{*} &=& \sum_{n=-\infty}^{\infty}\frac{l_{0}(n)}{\left[l_{0}^{2}(n)
+ L^{2} + x(1-x)q^{2}\right]^{2}}.
\end{eqnarray}

It is not easy to calculate Eq. (\ref{eq:pi_ij}) and Eq. (\ref{eq:pi_00}) analytically. Nevertheless, within the widely used instantaneous approximation $q_{0} = 0$, the integration can be performed by the methods presented in Ref. \cite{Dorey} and \cite{Liu09}. After tedious but straightforward computation, we obtain the following expressions for polarization functions:
\begin{eqnarray}
&&\Pi_{A}({\bf q},\beta) = \frac{2\alpha}{\pi\beta} \int_{0}^{1}dx\,
\frac{10^{\ln2}X_{\Gamma} + 1}{2}
\nonumber \\&&\,\,
\times\ln\left[2\left(\cosh
\left[\frac{2\pi X_{{\bf q}}(x)}{{10^{\ln2}X_{\Gamma} + 1}}\right] +
\cosh\left[\frac{2\pi
X_{\mu}}{10^{\ln2}X_{\Gamma}+1}\right]\right)\right],
\nonumber \\
&&\Pi_{B}({\bf q},\beta) = \frac{4\alpha}{\beta} \int_{0}^{1}dx\,
X_{{\bf q}}(x)
\nonumber \\&&\,\,
\times\frac{\sinh\left[\frac{2\pi X_{{\bf
q}}(x)}{10^{\ln2}X_{\Gamma} + 1}\right]} {\cosh\left[\frac{2\pi X_{{\bf
q}}(x)}{10^{\ln2}X_{\Gamma} + 1}\right] +
\cosh\left[\frac{2\pi X_{\mu}}{10^{\ln2}X_{\Gamma} + 1}\right]},
\end{eqnarray}
where $X_{{\bf q}}(x) = \frac{\beta}{2\pi}\sqrt{x(1 - x)}{\bf q}$, $X_{\Gamma} = \frac{\beta}{2\pi}\Gamma_0$, $X_{\mu} = \frac{\beta}{2\pi}\mu$. In either of the limits $\beta \rightarrow \infty$ ($T \rightarrow 0$) or ${\bf q} \rightarrow \infty$, both the above functions reduce to $\alpha{\bf q}/8$ and then the photon propagator reduces to the zero-temperature result (Eq. (\ref{eq:photon0})). However, in the limit ${\bf q} \rightarrow 0$, $\Pi_{A}({\bf q},\beta)$ becomes a function of temperature, chemical potential, and impurity scattering rate, $M_{00}(\beta, X_{\Gamma}, X_{\mu})$. In the zero energy and zero momentum limit, the photon propagator is
\begin{equation}
\Delta_{\mu\nu}(q_{0} = 0, {\bf q}\rightarrow 0, \beta) =
\frac{A_{\mu\nu}}{{\bf q}^{2} + M_{00}^{2}} + \frac{B_{\mu\nu}}{{\bf
q}^{2} + 0}.
\end{equation}
Apparently, the temporal component of the propagator now acquires an effective mass and becomes
\begin{equation}\label{eq:screen_length}
M_{00} =
\sqrt{\frac{2\alpha}{\pi\beta}(10^{\ln2}X_{\Gamma} + 1)
\ln\Big[2\cosh\big[\frac{\pi X_{\mu}}{10^{\ln2}X_{\Gamma} + 1}\big]\Big]}. \nonumber \\
\end{equation}
The mass appearing in the longitudinal photon implies that the electric field acquires a static screening length determined by the chemical potential and impurity scattering rate. The transverse photon, however, remains massless and hence the corresponding magnetic field is still long-ranged, albeit dynamically screened.

We now study the problem of DCSB at finite temperature. As in the case of zero temperature, this problem can be studied by the DS equation approach. The DS equation for fermion propagator $S_{F}$ at finite temperature is given by
\begin{eqnarray}\label{eq:ds_f}
&&S_{F}^{-1}(p_{0},\mathbf{p},\beta) = S_{F}^{(0) -1}(p_{0},\mathbf{p},\beta)
\nonumber \\&& \,\,-
\frac{e^{2}}{\beta}\sum_{n=-\infty}^{\infty} \int\frac{d^{2}k}{(2\pi)^{2}}
\gamma_{\mu}S_{F}(k_{0},\mathbf{k},\beta)\Gamma_{\nu}
\Delta_{\mu\nu}(q_{0},{\bf q},\beta),\nonumber \\
\end{eqnarray}
where $\Gamma_{\nu}$ is the fermion-photon vertex and $q = p - k$. The full fermion propagator can be written as
\begin{equation}\label{eq:fermion_full_t}
S_{F}^{-1}(p_{0},\mathbf{p},\beta) =
\mathcal{A}(p_{0},\mathbf{p},\beta)S_{F}^{(0) -1} + \Sigma(p_{0},\mathbf{p},\beta),
\end{equation}
where $\mathcal{A}(p_{0},\mathbf{p},\beta)$ is the wave-function renormalization. Substituting Eq. (\ref{eq:propagator_full}) and Eq. (\ref{eq:fermion_full_t}) into Eq. (\ref{eq:ds_f}) and then taking trace on both sides of this equation, we get the following couple of closed integral equations
\begin{eqnarray}\label{eq:dse}
&&\mathcal{A}(p_{0},\mathbf{p},\beta) =
1 +  \frac{e^{2}}{4\beta}\sum_{\omega_n}\int \frac{d^{2}\mathbf{k}}{(2\pi)^{2}}
\mathrm{Tr}\left[S_{F}^{(0)}\gamma_{\mu}S_{F}\Gamma_{\nu}
\Delta_{\mu\nu}\right],\nonumber \\
&&\Sigma(p_{0},\mathbf{p},\beta) =
- \frac{e^{2}}{4\beta}\sum_{\omega_n}\int \frac{d^{2}\mathbf{k}}{(2\pi)^{2}}
\mathrm{Tr}\left[\gamma_{\mu}S_{F}\Gamma_{\nu}
\Delta_{\mu\nu}\right].
\end{eqnarray}
Here, $\Delta_{\mu\nu}$ is the photon propagator as defined by Eq. (\ref{eq:photon_full_t}) and the $\Gamma_{\nu}$ is the full vertex function. To treat this equation, a number of approximations should be made.

At present, there is no well-controlled way to choose the vertex function $\Gamma_{\nu}$. In order to satisfy the Ward-Takahashi identity, Maris \emph{et} \emph{al.} \cite{Maris} studied several different \emph{Ans$\ddot{a}$tze} for the full vertex function and compared the results. For the sake of simplicity, they assumed that $\Gamma_{\nu} = f(\mathcal{A}(p),\mathcal{A}(k),\mathcal{A}(p - k)) \gamma_{\nu}$, with $\mathcal{A}(p) \equiv 1$ corresponding to the bare vertex. They found that, within a range of vertex functions, the critical behavior of DCSB is almost independent of the precise form of $f$-function and the bare vertex is actually a good approximation. Once $\Gamma_{\nu} = \gamma_{\nu}$ is assumed, then the Ward-Takahashi identity requires that the wave function renormalization should be $\mathcal{A}(p) \equiv 1$.

In the present case, it is much more difficult to solve the DS equations Eq. (\ref{eq:dse}) with finite chemical potential and impurity scattering than in the clean system. Unlike the zero temperature case, the particle energy and momentum should be treated separately in the finite temperature perturbation theory. Now, the energy becomes discrete frequencies proportional to the temperature $T$. To simplify the theoretical and numerical analysis, we keep only the leading order of $1/N$ expansion by replacing the vertex function $\Gamma_{\nu}$ by $\gamma_{\nu}$ and taking $\mathcal{A}(p)=1$. The DS equation of fermion Eq. (\ref{eq:fermion_full_t}) can now be simplified as
\begin{eqnarray}
S_{F}^{-1}(p_{0},\mathbf{p}, \beta) &=&
[(i\omega_{n} + i\Gamma_{0}\mathrm{sgn}(\omega_{n}) - \mu)\gamma_{0}
- \mathbf{\gamma}\cdot \mathbf{p}]
\nonumber \\&&
+ \Sigma(p_{0}, \mathbf{p}, \beta).
\end{eqnarray}
The fermion mass function $\Sigma(p_{0},\mathbf{p}, \beta)$ satisfies the following integral equation
\begin{eqnarray}\label{eq:dsf}
&& \Sigma(p_{0}, \mathbf{p}, \beta) = \frac{e^{2}}{\beta}\sum_{\omega_n}\int \frac{d^{2}\mathbf{k}}{(2\pi)^{2}} \mathrm{Tr}[\gamma_{\mu}\gamma_{\nu}\Delta_{\mu\nu}(q_{0},{\bf q},\beta)] \nonumber \\
&&\times\, \frac{\Sigma(k_{0}, \mathbf{k},\beta)}{ 4\left[(\omega_{n} + \Gamma_{0}sgn(\omega_{n}) + i\mu)^{2} + {\bf k}^{2} + \Sigma^{2}(k_{0}, \mathbf{k},\beta)\right]} \nonumber \\
&&= \frac{\alpha}{N}\frac{\beta}{4\pi^2}\sum_{n=-\infty}^{\infty}\int \frac{d^{2}\mathbf{k}}{(2\pi)^{2}}
\Delta_{\mu\mu}(q_{0},{\bf q},\beta) \nonumber \\
&&\times\, \frac{\Sigma(k_{0}, \mathbf{k},\beta)}{\left[((n + \frac{1}{2}) + X_{\Gamma}sgn(\omega_{n}) + iX_{\mu})^{2}+ M^{2}\right]},
\end{eqnarray}
where
\begin{eqnarray}
&&X_{T} = \frac{\beta}{2\pi},\,\,\,\,\,\,\
M = X_{T} \sqrt{{\bf k}^{2} + \Sigma^{2}(\mathbf{k},\beta)}.
\end{eqnarray}
Note that chemical potential and impurity scattering rate both appear in two places: in the occupation number and in the polarization function. The energy shift induced by chemical potential and the fermion damping effect induced by impurity scattering are both represented in the former place. The screening of gauge interaction caused by chemical potential and impurity scattering is reflected in the vacuum polarization function. In order to perform the frequency summation in Eq. (\ref{eq:dsf}), we utilize the approximation $\Sigma(p_{0}, \mathbf{p}, \beta) \simeq \Sigma(p_{0} = 0, \mathbf{p}, \beta)$. At $\omega_{m} = q_{0} = 0$, the summation over $\omega_{n}$ in Eq. (\ref{eq:dsf}) yields the equation
\begin{eqnarray}
\Sigma(\mathbf{p},\beta)
&=& \frac{\beta\alpha}{4N\pi^2}\int\frac{d^{2}\mathbf{k}}{(2\pi)^{2}}
\frac{\Sigma(\mathbf{k},\beta)}{M}\Delta_{\mu\mu}(q_{0} = 0,{\bf q},\beta)
\nonumber \\&&\times\Im m \left[\psi
\big[\frac{1}{2} + X_{\Gamma} \pm iX_{\mu} + i M\big]\right].
\end{eqnarray}
Substituting the full photon propagator Eq. (\ref{eq:photon_full}) into the mass equation and use the same approximative method as that presented in Ref. \cite{Liu09}, we have
\begin{eqnarray}\label{eq:mass}
&&\Sigma(\mathbf{p},\beta)
\nonumber \\
&=&\frac{\beta\alpha}{4N\pi^2}\int\frac{d^{2}\mathbf{k}}{(2\pi)^{2}}
\left[\frac{1}{q^{2} + \Pi_{A}(\mathbf{q}, \beta)} + \frac{1}{q^{2}
+ \Pi_{B}(\mathbf{q}, \beta)} \right]
\nonumber \\&&\times
\left\{\frac{\Sigma(\mathbf{k},\beta)}{M}\Im m \left[\psi
\big[\frac{1}{2} + X_{\Gamma} \pm iX_{\mu} + i M\big]\right]\right\}
\nonumber \\&\approx&
\frac{\beta\alpha}{4N\pi^2}\int\frac{d^{2}\mathbf{k}}{(2\pi)^{2}}
\left[\frac{1}{q^{2} + \Pi_{A}(\mathbf{q}, \beta)} + \frac{1}{q^{2}
+ \Pi_{B}(\mathbf{q}, \beta)} \right]
\nonumber \\&&\times
\left\{\frac{\Sigma(\mathbf{k},\beta)}{M}
\frac{\pi}{2}\tanh\left[\frac{\pi(M \pm
X_{\mu})}{10^{\ln2}X_{\Gamma} + 1}\right] \right\}.
\end{eqnarray}
If this nonlinear integral equation develops a nontrivial solution,
then the massless fermion acquires a finite mass.

\section{Solution of the Dyson-Schwinger equation}
\label{sec:solution}

The nonlinear integral equation can be solved by the bifurcation theory and parameter imbedding method \cite{Atkinson93,Cheng}. The basic idea and detailed computation procedures are presented in previous papers \cite{Atkinson93,Cheng, Liu03}. To determine the bifurcation point that separates the chiral symmetric phase and symmetry broken phase, we need to find the eigenvalues of the associated linearized equation. Taking the Fr\^{e}chet derivative of the nonlinear integral equation Eq. (\ref{eq:mass}), we have
\begin{eqnarray}\label{eq:mass_nc}
&&\Sigma(\mathbf{p},\beta)
\nonumber \\&=&
\frac{\beta\alpha}{4N\pi^2}\int\frac{d^{2}\mathbf{k}}{(2\pi)^{2}}
\left[\frac{1}{q^{2} + \Pi_{A}(\mathbf{q}, \beta)}
+ \frac{1}{q^{2} + \Pi_{B}(\mathbf{q}, \beta)}\right]
\nonumber \\&&\times
\Sigma(\mathbf{k},\beta)\frac{\partial}{\partial \Sigma}
\left\{\frac{\Sigma(\mathbf{k},\beta)}{M}
\frac{\pi}{2}\tanh\left[\frac{\pi(M \pm X_{\mu})}{10^{\ln2}X_{\Gamma} + 1}\right]
\right\}\Bigg|_{\Sigma=0}
\nonumber \\&=&
\frac{\beta}{4N\pi^2}\int\frac{d^{2}\mathbf{k}}{(2\pi)^{2}}
\left[\frac{1}{q^{2} + \Pi_{A}(\mathbf{q}, \beta)}
+ \frac{1}{q^{2} + \Pi_{B}(\mathbf{q}, \beta)}\right]
\nonumber \\ && \times
\frac{\Sigma(\mathbf{k},\beta)}{X_{T}|{\bf k}|}
\frac{\pi}{2}\tanh\left[\frac{\pi(X_{T}|{\bf k}| \pm X_{\mu})}{10^{\ln2}X_{\Gamma} + 1}\right].
\end{eqnarray}
To facilitate numerical calculation, we divide the momenta, temperature, fermion mass, chemical potential and impurity scattering rate by parameter $\alpha$ and make them dimensionless. The UV cutoff for momentum now is replaced by $\Lambda/\alpha$, which should be properly chosen to ensure the results insensitive to UV cutoff. As pointed out by Appelquist \emph{et} \emph{al.} \cite{Appelquist86, Appelquist88}, the integral in the zero temperature DS equation damps rapidly for momenta greater than $\alpha$, it is thus natural to set $\Lambda/\alpha \simeq 1$. This is also true for DS equation at finite temperature \cite{Aitchison92}. In the following we take the advantage of this fact and assume that $\Lambda/\alpha = 1$.

\begin{figure}[t]
  \centering
   \subfigure{
    \label{fig:subfig:NC_T_-8} 
    \includegraphics[width=2.7in]{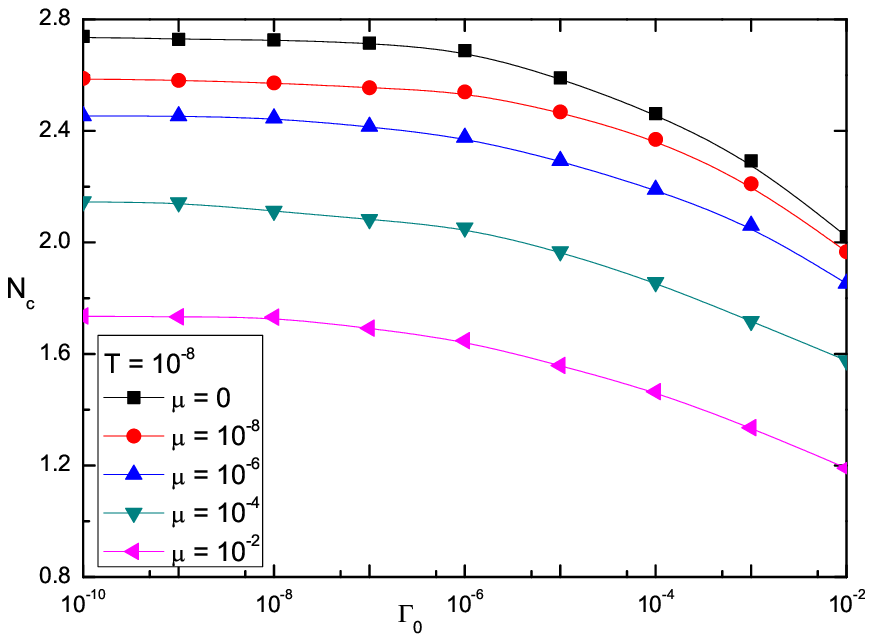}}
  \subfigure{
    \label{fig:subfig:NC_T_-6} 
    \includegraphics[width=2.7in]{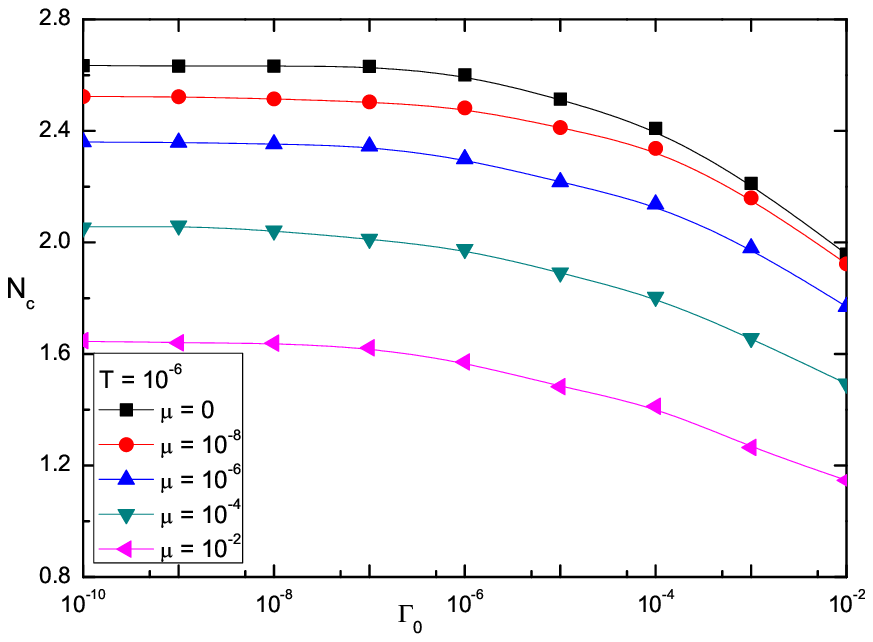}}
      \subfigure{
    \label{fig:subfig:NC_T_-4} 
    \includegraphics[width=2.7in]{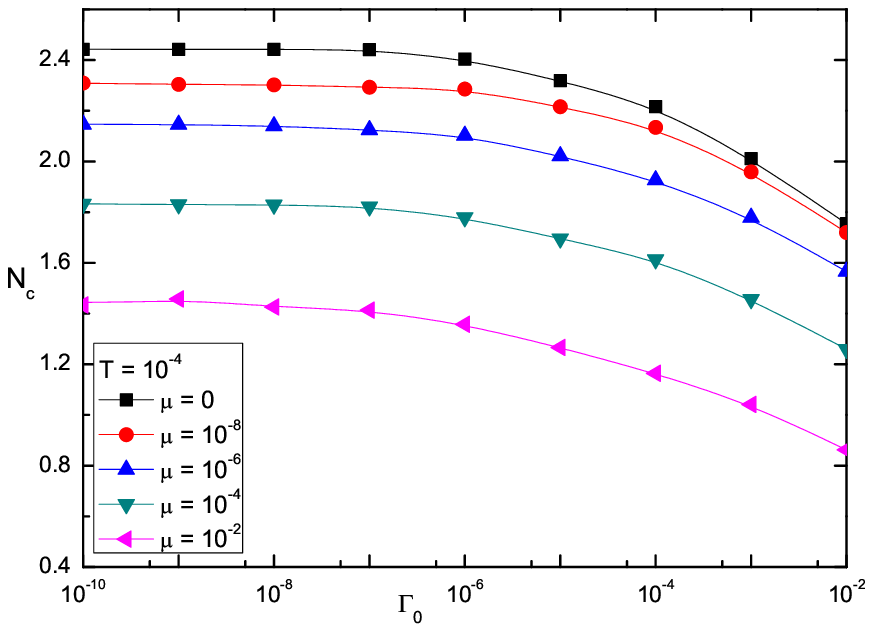}}
\caption{Dependence of $N_{c}$ on $\mu$ and $\Gamma_{0}$ for
different temperatures $T = 10^{-8}, 10^{-6}$ and $10^{-4}$.}
  \label{fig:NC} 
\end{figure}

The dependence of critical flavor number $N_c$ on chemical potential $\mu$ and impurity scattering rate $\Gamma_{0}$ are shown in Fig. \ref{fig:NC} for a list of values of temperature $T = 10^{-8}, 10^{-6}, 10^{-4}$. It is easy to see that the critical flavor $N_{c}$ is a decreasing function of $T$, $\mu$, and $\Gamma_{0}$, implying a strong suppression of DCSB by thermal fluctuation, fermion density, and impurity scattering.

\begin{figure}[t]
  \centering
    \includegraphics[width=2.7in]{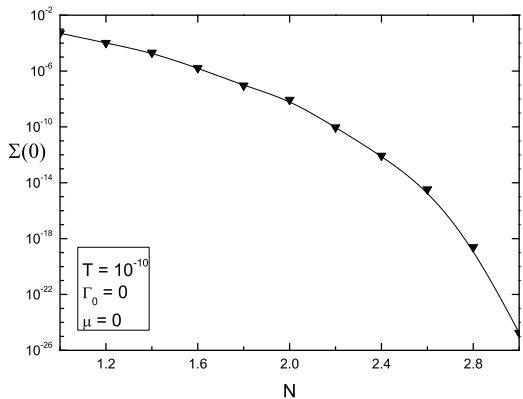}
\caption{Zero-momentum mass $\Sigma(0)$ for different values of N
at $T = 10^{-10}$ and $\mu = \Gamma_{0} = 0$.}
  \label{fig:mass_Nc} 
\end{figure}

\begin{figure}[t]
  \centering
   \subfigure{
    \label{fig:subfig:mass0_T_-8} 
    \includegraphics[width=2.7in]{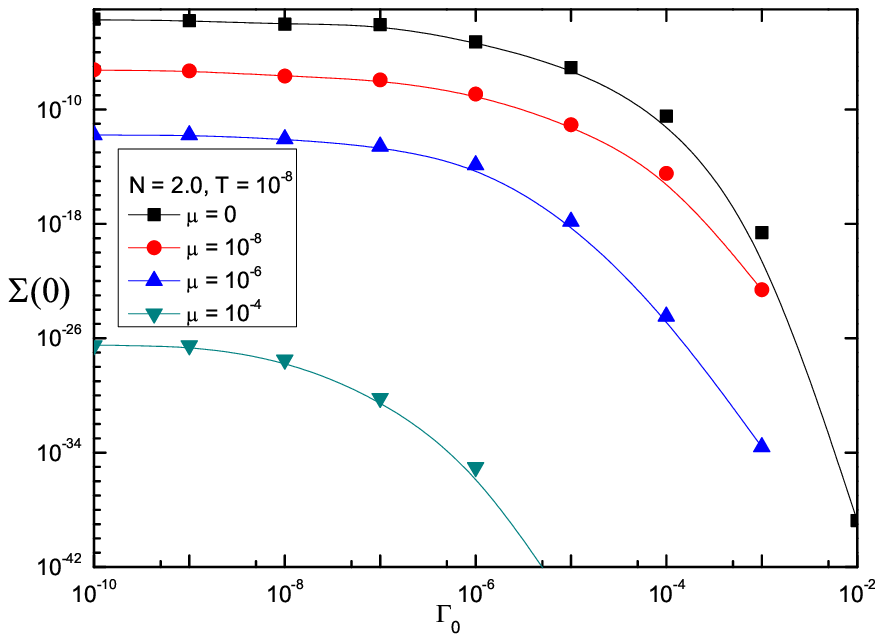}}
   \subfigure{
    \label{fig:subfig:mass_p} 
    \includegraphics[width=2.7in]{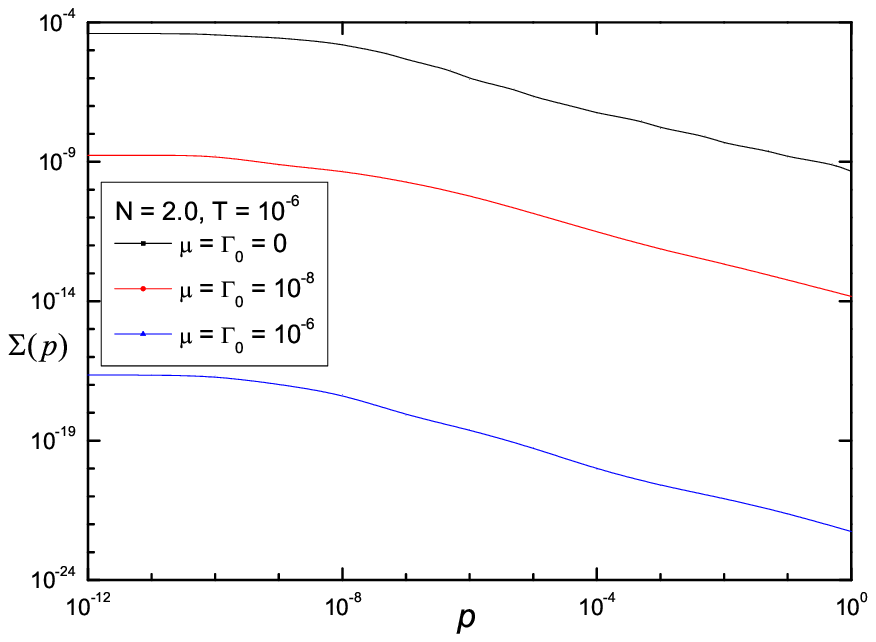}}
\caption{(a) Zero-momentum mass $\Sigma(0)$ for different values of
$\mu$ and $\Gamma_{0}$ at $T = 10^{-8}$;
(b) Dependence of mass $\Sigma(p)$ on momentum $p$ for different values of
$\mu$ and $\Gamma_{0}$ at $T = 10^{-6}$.}
  \label{fig:mass_p} 
\end{figure}

The fermion mass function $\Sigma(\mathbf{p},\beta)$ can be obtained after solving the DS equation Eq. (\ref{eq:mass}) using the straightforward iteration method. The results are presented in Fig. \ref{fig:mass_Nc} and Fig. \ref{fig:mass_p}. In Fig. \ref{fig:mass_Nc}, we show the re-scaled zero-momentum fermion mass $\Sigma(0)$ for several different values of fermion flavor $N$ at a fixed temperature $T = 10^{-10}$, with $\mu = \Gamma_{0} = 0$. Apparently, the fermion mass $\Sigma(0)$ decreases rapidly as the fermion flavor $N$ increases. This is easy to understand since the inverse of fermion flavor $1/N$ serves as the effective coupling strength of gauge interaction.

The dynamical fermion mass as a function of momentum is shown in Fig. \ref{fig:mass_p} for $N = 2$ at different values of temperatures, chemical potential and impurity scattering rate. We notice that the dynamical fermion mass is significantly suppressed by the increasing chemical potential and impurity scattering rate. It also decreases with the increasing fermion momentum.

We next would like to compare the results presented above with those in QED$_{3}$ with finite photon mass generated by Anderson-Higgs mechanism. Different from the partial screening of gauge interaction, the Anderson-Higgs mechanism induces a complete static screening. In this case, the gauge boson acquires a physical mass by eating the massless Goldstone bosons, so that the propagator (Eq. (\ref{eq:photon_full})) becomes
\begin{eqnarray}
\Delta_{\mu\nu}(q_{0}, {\bf q}, \beta) &=& \frac{A_{\mu\nu}}{q^{2} +
\Pi_{A}(q_{0},{\bf q},\beta) + m_{a}^{2}}
\nonumber \\ &&
+ \frac{B_{\mu\nu}}{q^{2}
+ \Pi_{B}(q_{0},{\bf q},\beta) + m_{a}^{2}}.
\end{eqnarray}
If $m_a \neq 0$, both temporal and spatial components of gauge field develop a finite screening length, which is given by the gauge boson mass. Such screening length eliminates the contribution of small momenta to the DS equation Eq. (\ref{eq:mass_nc}). It is expected that a large $m_a$ will prevent the DCSB. The dependence of critical flavor $N_c$ on $m_{a}$ and $\Gamma_{0}$ is shown in Fig. 4. The gauge boson mass and impurity scattering lead to similar suppression effect of DCSB.

\begin{figure}[ht]
  \centering
    \label{fig:subfig:Nc_Gamma_ma} 
    \includegraphics[width=2.7in]{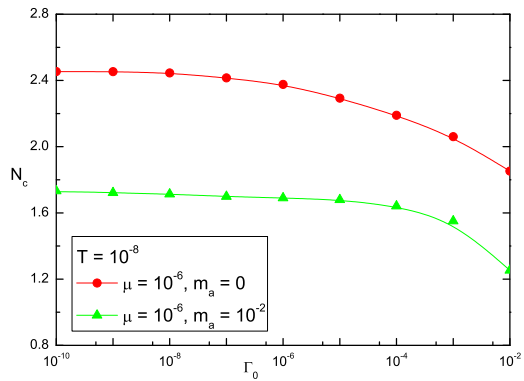}
\caption{Dependence of $N_c$ on $m_{a}$ and $\Gamma_{0}$ at $T = 10^{-8}$.}
  \label{fig:NCma} 
\end{figure}

\section{Summary and discussion}
 \label{sec:summary}

In summary, we studied the effects of finite chemical potential and
impurity scattering on dynamical mass generation in QED$_{3}$. In
the realistic applications of QED$_{3}$ to condensed matter systems,
these effects usually can not be ignored. By solving the DS equation
for fermion mass function, we found that both chemical potential and
impurity scattering lead to strong reduction of critical fermion
flavor $N_{c}$ and dynamical fermion mass. In reality, the chemical
potential, impurity scattering, and even a finite gauge boson mass
may coexist at the same time. When their effects are all important,
the DCSB is completely suppressed. These results impose an important
constraint on the applicability of QED$_{3}$ in various physical
systems. For a system with large fermion density and impurity
potential, it seems impossible that DCSB can indeed take place.

We now would like to briefly discuss the meaning of finite chemical
potential and the possible experimental study of its effect on DCSB.
In condensed matter physics, the zero temperature ground state
(vacuum) of a many-body system is characterized by the presence of a
Fermi level which separates the fully occupied and fully empty
states. In some condensed matter systems, such as high temperature
superconductor and graphene, the valence band and conduction band
touch at discrete Dirac points. When the Fermi level lies exactly at
Dirac points, the chemical potential is zero and the low-energy
excitations are massless Dirac fermions. The Fermi level moves
upwards (downwards) from the Dirac points once the fermion density
increases (decreases). Now the chemical potential is just the
quantity that measures how the new Fermi level is far from the
original Dirac points. At zero chemical potential, the pairs are
formed by fermions slightly above Dirac points and anti-fermions
(holes in condensed matter terminology) slightly below the Dirac
points. Once the fermions become massive, they are confined. At
finite chemical potential, the pairs are formed by fermions slightly
above the finite Fermi surface and anti-fermions slightly below
Fermi surface. The massive fermions are also confined by the gauge
force.

In realistic condensed matter systems, the fermion density or
chemical potential can be continually turned by chemical doping or
gate voltage. We first have a number of samples, each with different
chemical potential, and let them stay at some finite temperature
where DCSB is completely prevented by thermal fluctuations. We then
cool down these samples and study whether DCSB indeed take place at
very low temperature by measuring some observable quantities, such
as specific heat, susceptibility, and electric conductivity. Through
this way, in principle the effect of finite chemical potential on
the fate of DCSB can be experimentally studied.

We emphasize that our treatment on the random potential applies only
to the case of weak impurity potential where the self-consistent
Born approximation is valid. When the random potential caused by
impurities takes the form of Gaussian white noise, such treatment
might be questionable. For this kind of impurities, the random
potential has to be averaged by performing proper functional
integration, which yields an effective four-fermion interaction in
the whole action. This new effective interaction can contribute a
new term to the DS equation. At present, it is unclear technically
how to study the DCSB and the fermion damping effect in a unified
framework. This problem is subject to further investigation.

\section{Acknowledgments}

We would like to thank G. Cheng, H. Feng, and J.-R. Wang for very
helpful discussions. This work is supported by the National Science
Foundation of China under Grant No. 10674122.

\end{document}